\documentclass{article}

\usepackage{PRIMEarxiv}
\usepackage{multirow}
\usepackage{pgfplots}
\usetikzlibrary{spy,backgrounds}
\usepackage[utf8]{inputenc} 
\usepackage[T1]{fontenc}    
\usepackage{hyperref}       
\usepackage{url}            
\usepackage{booktabs}       
\usepackage{amsfonts}       
\usepackage{nicefrac}       
\usepackage{microtype}      
\usepackage{lipsum}
\usepackage{fancyhdr}       
\usepackage{graphicx}       
\graphicspath{{media/}}     

\title{Enhancement of Power Grid Monitoring Based on Data Weighting
}

\author{
  Parisa Ataeian\thanks{Corresponding Author} , Abbas Rabiee \\
  Department of Electrical Engineering, University of Zanjan , Zanjan , Iran  \\
  \texttt{\{ataeian.parisa, rabiee\}@znu.ac.ir} \\
   \And
  Mehdi Derafshian Maram, Mohsen Ghalei Monfared Zanjani \\
  Iran Grid Management Company, Tehran, Iran \\
  \texttt{\{derafshian, monfared.m\}@imc.ir} \\

}

\begin{document}
\maketitle

\begin{abstract}
With their expansion, national power grid have had to work with huge sets of data received from a vast number of substations and power plants. Given their large volume and variety, these data can be classified as big data. Managing this massive amount of data is certainly challenging. Depending on the application, parts of these data are more important for real-time network operation. Computing a network’s observability score without assigning weights to different signals may not provide a complete picture of the received data’s validity and thus lead to incorrect assessments of the network status. Consequently, signals critical to the network operation and functions of an energy management system (EMS) should be assigned higher weights in observability calculations. The weighted observability alongside the classic non-weighted observability can serve as an indicator of each area’s condition in comparison to that of other areas and so greatly facilitate the monitoring and verification of transmitted data. For calculating a weighted observability, the current paper presents a method based on the Analytic Hierarchy Process (AHP), in which higher weights are assigned to data that are more valuable for and widely used by operators. The national electricity network of Iran was chosen for the evaluation of the proposed method’s effect on data quality and operational risk. The results indicated that the introduced method positively affects the quality of received data and also corrects erroneous data in the network.

\end{abstract}

\keywords{Observability \and 
Operation quality \and Information weighting \and Information quality \and Big data \and AHP}

\section{Introduction}
With the expansion of national power grids during the past decades, there has been a substantial rise in the amount of data received from stations as well as a growing dependence on these data for controlling power systems. Due to their large volume and variety, these data can be classified as big data \cite{daki2017big}. In regard to working with such big data, the volume, variety, and retrieval speed of the data are crucial. Furthermore, the accuracy of these data cannot be overlooked, because incorrect information may cause operator errors \cite{handschin1975bad}. The lack of access to reliable information was a major cause of large blackouts in North America and parts of southern Canada in 2003 as well as in Europe in 2006 \cite{hauer2004performance,andersson2005causes}. 
\\
For a power network to operate safely, it is necessary that all network variables always be within the allowable range and that the system be able to maintain its normal operation after a wide variety of potential events. These events can be predicted with expert tools, such as an energy management system (EMS) with application modules, an example of which is state estimation. Depending on the conditions of the power grid, data collection and data transfer to the control center may encounter various problems. Solving or at least alleviating these problems can greatly contribute to the quality, safety, and security of the network operation \cite{wu2005power}.
\\
The collection, transmission, and monitoring of data from power plants and substations are usually performed with two systems: WAMS and SCADA. WAMS offers excellent features, such as data synchronization, linearity of equations, dynamic analyzability, and the need for less equipment that results in a lower failure risk. However, due to its high costs and lack of a suitable telecommunications infrastructure, WAMS cannot completely replace SCADA \cite{phadke1993synchronized}. Therefore,  the usage of both systems is common \cite{li2008recognizing,anjia2005pmu}. The present article focuses on SCADA.
\\
Among all the different kinds of data received in dispatching centers, those that are widely used in the control center are a priority \cite{morrow2012topology,merrill1971bad}. Any interference with proper data collection and retrieval (e.g. delays in retrieval, poor condition of equipment, and incorrect measurements) can seriously disrupt the state estimation process \cite{kotiuga1982bad,hongxun2018data,falcao1982power}. The “observability” of a station (power plant or substation) or area indicates what percentage of all data sent to the control center is correct. In observability estimations, it is common to consider all of the data sent from a station or area. However, the presence of data that appear to be true, but are in fact false, can cause errors in observability estimations which may subsequently cause operator mistakes. Such apparently valid but actually invalid data can be identified through the state estimation process \cite{monticelli1983reliable,peterson1988multiple,wu1988observability}.
\\
In the rest of the present article, Section 2 describes the method for identifying invalid data and classifying operationally important data. Section 3 provides an introduction to the Analytic Hierarchy Process (AHP) and explains how it is used in the current study. Section 4 presents the results of applying weighting coefficients to observability and Section 5 concludes the paper.

\section{Identification and classification of important data from the operators’ perspective}
\label{sec:headings}
\subsection{Importance of data from the operators’ perspective}
In the online monitoring and control of power networks, up-to-date data from network components are crucial in addition to knowing the correct status of the operating instructions. Given the massive size of national networks, it is absolutely necessary for operators to determine the status of network operating points through the state estimation module \cite{morrow2012topology}.
To estimate the real-time operating point of the power system, SCADA’s state estimation module employs the raw measurement data collected from SCADA subsystems, including the voltage size and angle of all network buses \cite{abur1990bad,huang2012state}. In order to obtain the right output from this module, appropriate inputs must be given, including network topology (presence or absence of components), the generating power of the units, the voltage of certain buses, the load of certain transformers, the load of capacitors and reactors, and the tap of transmission transformers. Obviously, the accuracy of the state estimation module’s output largely depends on the quality of these inputs and measurements \cite{li2008recognizing,savulescu2009real,mazhabjafari2010feasibility}.
\subsection{Data tagging}
Table \ref{signaltag} provides a list of potential status tags of data received in the control center. The received data may be either valid or invalid for various reasons.
\begin{table}[!h]
\centering
\caption{Data tags}
\label{signaltag}
\begin{tabular}{lc}
\toprule
\textbf{Definition}   & \textbf{Signal Tag} \\ \hline\hline
No information on the data status   & Faulty (F) \\ \hline\hline
Data show the last status before disconnection    & Non-current (N) \\ \hline
Valid data (these data are not tagged)       & Valid (V)    \\ \hline
Data appear to be valid but are invalid (these data are not tagged) & Valid (V)    \\ 
\hline
Data are made invalid so as to be disregarded in the state estimation process & Invalid (I)    \\ 
\hline
Important data that are sometimes manually entered by operators.\\ Because of the constant changes in the network operating point, \\these data must obviously be constantly updated. & Manually (M)    \\ 
\bottomrule
\end{tabular}
\end{table}
\\
Apparently valid but inherently invalid data can lead to poor operating performance and decisions and, unfortunately, these data  cannot be easily identified as they have no specific tag or label. In the state estimation process, one of SCADA’s functions is to identify data points when there is a significant difference between estimated values and measurements. 
\subsection{Selecting important data}
The received data are selected from among the data that are important for different parts of the network operation, that is, according to the needs assessment and experiences of operators as well as the inputs required by the state estimation module. Table \ref{table2} presents the classification of signals by the type of quantity.
\begin{table}[!h]
\centering
\caption{Classification of signals based on information type}
\label{table2}
\begin{tabular}{lc}
\toprule
\textbf{Data} & \textbf{Tags} \\ \hline\hline
Megawatt & MW \\ \hline\hline
Mega Volt*Amps Reactive & MV \\ \hline
Kilovolt & KV    \\ \hline
TAP & TAP   \\ \hline
Status & STATUS   \\
\bottomrule
\end{tabular}
\end{table}
\begin{table}[!h]
\centering
\caption{Classification of signals by the type of component}
\label{table3}
\begin{tabular}{lc}
\toprule
\textbf{Signals} & \textbf{Components} \\ \hline\hline
1 & Unit Transformer and Load Transformer \\ \hline\hline
2 & Transmission Transformer \\ \hline
3 & Generator    \\ \hline
4 & Transmission Line   \\ \hline
5 & Reactor,  Capacitor  \\ \hline
6 & Busbar   \\
\bottomrule
\end{tabular}
\end{table}
\\
In this classification, signal Output MW, for example, has the same value as signal Line MW. Therefore, these signals should be differentiated. Table \ref{table3} shows the differentiation of signals by the type of component.
In order to weight the signals, the set of received data must be examined from two perspectives: 1) type of quantity (unit) and 2) the component they come from. Since not all components use all the quantities listed in Table \ref{table2}, the quantity related to each component is considered, as shown in Table \ref{table4}.
\begin{table}[!h]
\centering
\caption{Related quantity to each component}
\label{table4}
\begin{tabular}{lc}
\toprule
\textbf{Component} & \textbf{Related quantity} \\ \hline\hline
Unit Transformer and Load Transformer & MW, MV, TAP, and STATUS\\ \hline
Transmission Transformer & MW, MV, TAP, and STATUS\\ \hline
Generator &   MW, MV, KV, and STATUS\\ \hline
Transmission Line & MW, MV, KV, and STATUS\\ \hline
Reactor,  Capacitor & MV and STATUS\\ \hline
Busbar &  KV and STATUS\\
\bottomrule
\end{tabular}
\end{table}
\section{Proposed formulation and algorithm}
\subsection{Observability}
The classic definition of observability does not differentiate the received data in terms of their importance. Equation \ref{observability} provides the formula of the observability index.
\begin{equation}
\label{observability}
    OV_{i} = \frac{AD_{i}-OB_{i}}{AD_{i}},
\end{equation}
where ${AD_{i}}$ is the total infotmation that is sent by state ${i}$ to control center and ${OB_{i}}$ is incorrect data. Table \ref{table5} presents another problem that calculates observability while giving the same weight to two stations  (transmission substation and a power plant station).
\begin{table}[!h]
\centering
\caption{Difference between two stations with the same observability index}
\label{table5}
\begin{tabular}{clccc}
\toprule
\textbf{Station} & \textbf{Uncorrect signals}& \textbf{\#Uncorrect signals}  & \textbf{\#Total signals}& \textbf{Observability indicator}\\ \hline\hline
A& MW of unit and breaker of head line reactor & 2&100&\%98 \\ \hline
B& Bus frequency and kV of line & 2&100&\%98 \\
\bottomrule
\end{tabular}
\end{table}
\\
While the two stations have the same observability according to the control center’s experts, Station B is in better condition because MW of unit and shunt reactor condition are more important than bus frequency and kV of line.
\subsection{Proposed weighting approach}
To solve the above problems, the current paper proposes weighting the data according to their importance to the network operation and the state estimation process. The present study conducts this weighting through a procedure based on Analytic Hierarchy Process (AHP).
\subsection{Analytic Hierarchy Process (AHP)}
Individuals may make decisions without considering all aspects of the issue at hand or when influenced by others or their own personal background. It is, therefore, advisable that the decision-making process employ methods which are more accurate, comprehensive, and rational. One of the most commonly used is the Analytic Hierarchy Process (AHP).
In AHP, available options are classified according to several criteria and decisions are made by comparing pairs of options in terms of each criterion \cite{golden1989analytic,saaty1989group}. With the assistance of several experts in the field, the present study utilized AHP to determine the operational importance of each datum. Since not all components have all of the considered quantities, AHP was applied separately for each component and each quantity. After obtaining the evaluations of each expert, the current work produced the final evaluations by averaging. The obtained averages were considered as the basis for determining the weight assigned to the signals.
\subsection{Questionnaire}
A questionnaire was prepared for the pairwise comparison of signals. Considering the number of operators in the studied network, the present study provided its questionnaire to nine experienced operators as well as experts in network operations and state estimation modules. Also examined was the issue of real-time operation control from the perspectives of the frequency control, voltage control, and reactive power of the network.
\section{Results and analysis}
For its evaluation, the proposed weighted observability method was applied to the Iranian electricity network. The network consists of ten areas, from A to J, each of which may be more or less critical depending on geographical factors (important exchange points), political factors, economic factors (industry), and the size of the network (the number of stations and signals). The current study assigned the same importance to all of these areas. Since the signals of critical regional stations are usually considered in regional exchange line operating instructions, these signals were given higher weights than other signals.
\subsection{Signal weighting based on expert inputs}
The results of each expert completed questionnaire were imported into the AHP software to determine the signal weights recommended by that expert. For example, Table \ref{table6} reports an expert’s opinion on the importance of signals received from a generation unit from the viewpoint of the state estimation module. 
\begin{table}[!h]
\centering
\caption{Expert weights on components}
\label{table6}
\begin{tabular}{lc}
\toprule
\textbf{Measure} & \textbf{Weight} \\ \hline\hline
MW & 41\% \\ \hline
MV & 23.3\% \\ \hline
KV &  40\% \\ \hline
Status & 31.7\%\\ 
\bottomrule
\end{tabular}
\end{table}
\\
For this component, the examined signals are MW, MVAR, kV, and connected/disconnected status. As seen, the unit’s MW signal was more important than its kV for the state estimation module. The connected/disconnected status of the unit and its MVAR were not as important. Obviously, the expert may not have given the same ratings to these signals because of other reasons. Similar pairwise comparisons were also made between the components from the quantity point of view. After collecting the results of the questionnaires and importing the data into the software, the current work averaged the resulting ratings, the results of which are listed in Tables \ref{table7} and \ref{table8}. In the averaging of the ratings, all of the expert inputs were considered to be of equal importance. The product of values given in Tables \ref{table7} and \ref{table8} was considered as the final weight of operationally important signals. Table \ref{table2}, \ref{table3} and \ref{table4} reports the importance of each quantity for each component. For example, the most important quantity for a transmission transformer was MW, followed by MVAR, TAP, and Status. As seen, because of the normalization of the geometric means’ results, the weights in each column summed up to 100.
\begin{table}[!h]
\centering
\caption{Resulting rating based on information type}
\label{table7}
\resizebox{\columnwidth}{!}{\begin{tabular}{ccccccc}
\toprule
\textbf{M} & \textbf{Unit Transformer and Load Transformer} & \textbf{Transmission Transformer}  & \textbf{Generator}& \textbf{Transmission Line} & \textbf{Reactor,  Capacitor} & \textbf{Busbar} \\ \hline\hline
MW& 57.3 & 45.07 & 51.5  & 59.97& - & -  \\ \hline
MV& 17.63 & 21.8 & 23.27 & 24.53 & 61.1 & - \\ \hline
KV& - & - & 7.37 & 6.27 & - & 71.1 \\ \hline
TAP& 4.87 & 18.37 & -& -& -&- \\ \hline
STATUS& 19.83& 14.8&17.9 &9.27 & 38.87&28.9 \\ \hline
\bottomrule
\end{tabular}}
\end{table}

\begin{table}[!h]
\centering
\caption{Resulting rating based on type of component}
\label{table8}
\begin{tabular}{cccccc}
\toprule
\textbf{N} & \textbf{MW} & \textbf{MV}  & \textbf{KV}& \textbf{TAP} & \textbf{STATUS} \\ \hline\hline
Unit Transformer and Load Transformer& 13.07 & 8.53 & -  & 32.2& 12.03  \\ \hline
Transmission Transformer& 15.33 & 19.43 & - & 67.77 & 16.93 \\ \hline
Generator& 54.57 & 49.97 & 39.93 & - & 40.97  \\ \hline
Transmission Line& 17.03 & 8.6 & 14.53 & - & 16.5 \\ \hline
Reactor,  Capacitor& -& 13.47 & - & - &9.07 \\ \hline
Busbar,  Capacitor& - & - & 45.53  & - &4.6 \\ \hline
\bottomrule
\end{tabular}
\end{table}
Table \ref{table2}, \ref{table3} and \ref{table4} provides the importance of each component from the perspective of each quantity. To simultaneously consider the importance of components and quantities, the weight of each signal in the observability calculations was defined as $M\times N$. For example, the weight of kV for the transmission lines in all stations was determined as $14.53 \times 6.27=91.10$. After calculating all weights in the same way, the present study observed that the kV of the busbar and the MW of the generator were the most important signals. As the voltage levels of all substations were displayed on a dedicated screen, it was obvious that these signals were important for operation. The megawatt output of the units was also critical for frequency control. The signal with the lowest importance was found to be the kV of the transmission lines; this was predictable given the redundancy of this data in each station.
\subsection{Observability index}
After calculating all the signal weights, the present study obtained the weighted observability by inserting the final weights in Equation \ref{eq2}: 
\begin{equation}
    \label{eq2}
    OV_{i}^{new}=\frac{AD_{i}^{new}-OB_{i}^{new}}{AD_{i}^{new}}*100,
\end{equation}
where ${OV_{i}^{new}}$ is the observability indicator and ${AD_{i}^{new}}$ calculated as follows: 
\begin{equation}
    AD_{i}^{new} = M_{i}*N_{i}*AD_{i}*2
\end{equation}
In Equation \ref{eq2}, ${AD_{i}^{new}}$ sum of all weighted signals (obtained from Equation 3).
Also, ${OB_{i}^{new}}$ new in Equation 2 is the sum of all weighted invalid data:
\begin{equation}
    OB_{i}^{new}=M_{i}*N_{i}*OB_{i}*2
\end{equation}
The multiplier of 2 is applied when the signal is present in the instruction so as to ensure that its importance is duly noted. During the calculation of the weighted observability index, it should be noted that data considered unimportant from an operating perspective are disregarded. Aside from this reason, such signals are also discounted in part because of their large number which makes pairwise comparisons impossible. 
\subsection{Observability results for the Iran power grid}
Both weighted observability and non-weighted observability were calculated for the Iran power grid, which consists of 16 areas. The following compares these two observability values to demonstrate the effect of differentiating important signals. Table \ref{table9} presents the observability of the considered network with and without the weight coefficients. It can be seen that, in the non-weighted approach, area A ranked first and area P ranked last in terms of observability.
In any case, computing observability without weighting the signals is impractical and does not meet the needs of operators in the safety area. In the current work, applying the weights to observability significantly decreased the observability of areas N, M, L, J, D, and O. In other words, the non-weighted observability of these areas was not useful from an operational point of view and many of their invalid data were actually important data. In contrast, there was no change in area B’s observability, which indicates that the non-weighted observability of this area was adequate from an operational perspective. Applying the weights also changed the ranking of the areas. As shown in Table \ref{table9}, Areas C and D both had a non-weighted observability of 97\%, while Area C had more favorable conditions from an operational standpoint. Illustrated in Figure 1, the results of Table \ref{table9} show a significant decline in the observability of many areas after the weights were applied to the observability formula.
\begin{table}[!h]
\centering
\caption{Network observability including weighting coefficients}
\label{table9}
\begin{tabular}{cccc}
\toprule
\multicolumn{2}{c}{\textbf{Without weighting}} & \multicolumn{2}{c}{\textbf{With weighting}} \\ \hline \hline
    Area      &    Information quality indicator      &         Area  &     Information quality indicator     \\ \hline
  A        &    \%99      &     B     &     \%98      \\ \hline
  B        &    \%98      &     A     &     \%97      \\ \hline
  C        &    \%97      &     C     &     \%95      \\ \hline
  D        &    \%97      &     F     &     \%90      \\ \hline
  E        &    \%96      &     K     &     \%90      \\ \hline
  F        &    \%95      &     E     &     \%89      \\ \hline
  G        &    \%95      &     H     &     \%89      \\ \hline
  H        &    \%93      &     G     &     \%88      \\ \hline
  I        &    \%92      &     D     &     \%87      \\ \hline
  J        &    \%90      &     I     &     \%86      \\ \hline
  K        &    \%88      &     J     &     \%80      \\ \hline
  L        &    \%88      &     L     &     \%79      \\ \hline
  M        &    \%87      &     P     &     \%75      \\ \hline
  N        &    \%86      &     M     &     \%74      \\ \hline
  O        &    \%84      &     N     &     \%74      \\ \hline
  P        &    \%80      &     O     &     \%68      \\ \hline
          \bottomrule
\end{tabular}
\end{table}
\subsection{Analysis of 3-month observability}
The weighted and non-weighted observability results were sent to the areas with invalid signals and, after three months, the observability of these areas was reexamined. Table \ref{table10} reports the areas’ weighted and non-weighted observability indexes after this period. As the results demonstrate, most areas showed significant growth in observability.
As illustrated in Figure 2, Areas F, B, and L experienced decreased observability. This may be related to long periods of disconnection in these areas because of some maneuvers occurring in those areas. observability remained unchanged in Areas H, D, C, and A, while it improved in other areas. Since this diagram alone cannot indicate changes in important data (signals), the changes in weighted observability are plotted in Figure 3.
Figure 3 shows that, except for Areas B and H, where observability decreased, and areas F and L, where it remained unchanged, the rest of the areas earned better observability scores than they had three months earlier.
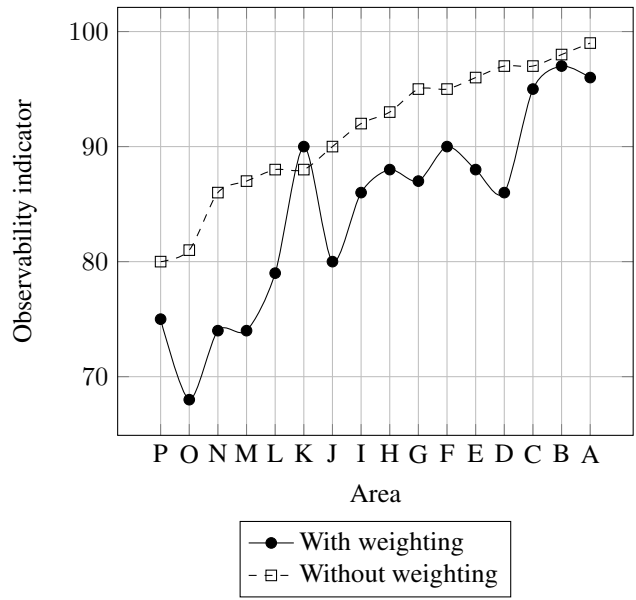
\begin{figure}[h]
\belowcaptionskip = -10pt
\centering
\begin{tikzpicture}
    \begin{axis}
        [grid = major,
         ylabel = {Observability indicator},
         xlabel = {Area},
         xtick=data,
         symbolic x coords={P,O,N,M,L,K,J,I,H,G,F,E,D,C,B,A},
         legend cell align={left},
         legend style={at={(0.5,-0.2)},anchor=north},
		legend entries={With weighting, Without weighting},
        ]
        \addplot [ color=black,
                solid,
                mark=*,
                mark options={solid},
                smooth ] coordinates
        {(P,75) (O,68) (N,74) (M,74) (L,79) (K,90) (J,80) (I,86) (H,88) (G,87) (F,90) (E,88) (D,86) (C,95) (B,97) (A,96)};
                \addplot [                color=black,
                dashed,
                mark=square,
                mark options={solid},
                smooth] coordinates
        {(P,80) (O,81) (N,86) (M,87) (L,88) (K,88) (J,90) (I,92) (H,93) (G,95) (F,95) (E,96) (D,97) (C,97) (B,98) (A,99)};
    \end{axis}
\end{tikzpicture}
\caption{Comparision of weight and non-weight observability}
\label{fig1}
\end{figure}
\begin{figure}[h]
\belowcaptionskip = -10pt
\centering
\begin{tikzpicture}
    \begin{axis}
        [grid = major,
         ylabel = {Observability indicator},
         xlabel = {Area},
         xtick=data,
         symbolic x coords={P,O,N,M,L,K,J,I,H,G,F,E,D,C,B,A},
         legend cell align={left},
         legend style={at={(0.5,-0.2)},anchor=north},
		legend entries={First observibility, Observibility (after 3 months)},
        ]
        \addplot [ color=black,
                solid,
                mark=*,
                mark options={solid},
                smooth ] coordinates
        {(P,80) (O,81) (N,86) (M,87) (L,88) (K,88) (J,90) (I,92) (H,93) (G,95) (F,96) (E,97) (D,97) (C,98) (B,98) (A,99)};
                \addplot [                color=black,
                dashed,
                mark=square,
                mark options={solid},
                smooth] coordinates
        {(P,85) (O,82) (N,88) (M,88) (L,86) (K,90) (J,93) (I,93) (H,94) (G,97) (F,97) (E,97) (D,97) (C,97) (B,97) (A,99)};
    \end{axis}
\end{tikzpicture}
\caption{Comparision of two weightless observability after three months}
\label{fig1}
\end{figure}
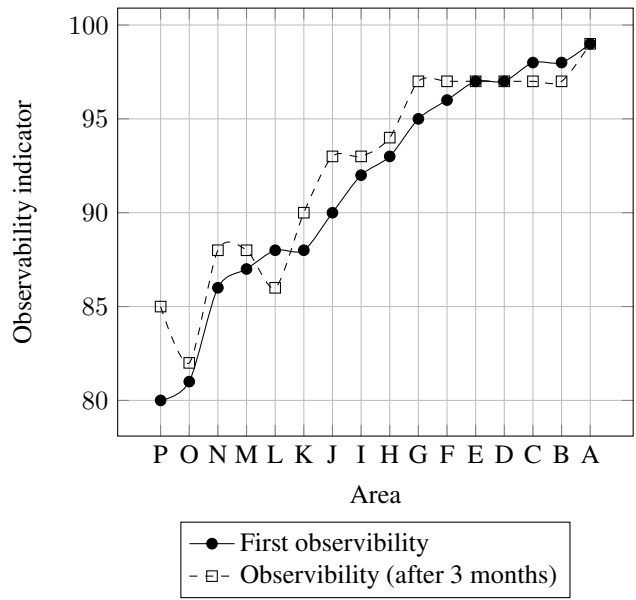
\vspace{1cm}
\begin{figure}[h]
\belowcaptionskip = -10pt
\centering
\begin{tikzpicture}
    \begin{axis}
        [grid = major,
         ylabel = {Observability indicator},
         xlabel = {Area},
         xtick=data,
         symbolic x coords={P,O,N,M,L,K,J,I,H,G,F,E,D,C,B,A},
         legend cell align={left},
         legend style={at = {(0.5,-0.2)},anchor=north},
		legend entries={First observibility, Observibility (after 3 months)},
        ]
        \addplot [ color=black,
                solid,
                mark=*,
                mark options={solid},smooth ] coordinates
        {(P,75) (O,68) (N,74) (M,74) (L,79) (K,90) (J,80) (I,86) (H,88) (G,87) (F,90) (E,88) (D,86) (C,95) (B,97) (A,96)};
                \addplot [                color=black,
                dashed,
                mark=square,
                mark options={solid},
                smooth] coordinates
        {(P,77) (O,79) (N,79) (M,78) (L,79) (K,92) (J,85) (I,89) (H,88) (G,91) (F,90) (E,95) (D,90) (C,96) (B,97) (A,98)};
    \end{axis}
\end{tikzpicture}
\caption{Comparision of two weights observability after three months}
\label{fig1}
\end{figure}
\begin{table}[!h]
\centering
\caption{Observability of the network (after 3 months)}
\label{table10}
\begin{tabular}{cccc}
\toprule
\multicolumn{2}{c}{\textbf{Without weighting}} & \multicolumn{2}{c}{\textbf{With weighting}} \\ \hline \hline
    Area      &    Information quality indicator      &         Area  &     Information quality indicator     \\ \hline
  A        &    \%99      &     A     &     \%98      \\ \hline
  B        &    \%97      &     B     &     \%97      \\ \hline
  C        &    \%97      &     C     &     \%96      \\ \hline
  D        &    \%97      &     E     &     \%95      \\ \hline
  E        &    \%97      &     K     &     \%92      \\ \hline
  G        &    \%96      &     G     &     \%91      \\ \hline
  F        &    \%94      &     F     &     \%90      \\ \hline
  H        &    \%93      &     D     &     \%89      \\ \hline
  I        &    \%93      &     I     &     \%89      \\ \hline
  J        &    \%90      &     H     &     \%88      \\ \hline
  K        &    \%90      &     J     &     \%84      \\ \hline
  M        &    \%88      &     L     &     \%79      \\ \hline
  N        &    \%88      &     N     &     \%79      \\ \hline
  L        &    \%86      &     O     &     \%79      \\ \hline
  P        &    \%85      &     M     &     \%78      \\ \hline
  O        &    \%82      &     P     &     \%78      \\ \hline
          \bottomrule
\end{tabular}
\end{table}
\section{Conclusion}
The present paper presents a weighted observability calculation method which places more emphasis on receiving valid data for signals considered more critical by operators and the state estimation module. To identify these important signals, nine experts in real-time operation control were asked to rate signals while taking into account the inputs required in the state estimation module. The experts completed questionnaires prepared for the pairwise comparisons of important signals, the results of which were analyzed by AHP software for determining each expert’s signal weights. The final signal weights were obtained by averaging these individual weights and then were applied to the observability formula. With the performance of these steps, periodic reports of important invalid signals can be sent to their origins so that the data may be corrected. The result of this process will be the continuous correction of erroneous data that produces a network with high observability and more reliability from an operator’s point of view. Furthermore, while the classic non-weighted approach to observability ignores the possibility of apparently valid but actually invalid data, the weighted approach ensures that such data are properly considered in observability calculations. Another advantage of the proposed approach is its attention to correcting certain data over others. It is crucial to assign a higher priority to more important data, especially in DCS substations where data transmission is prioritized. By improving network observability, these modifications can reduce faults and errors due to faulty operator performance. Providing more accurate input data to the state estimation module also lowers the convergence error and accelerates the process of real-time assessment and control.

\bibliographystyle{unsrt}  
\bibliography{references}

\begin{thebibliography}{10}

\bibitem{daki2017big}
Houda Daki, Asmaa El~Hannani, Abdelhak Aqqal, Abdelfattah Haidine, and Aziz
  Dahbi.
\newblock Big data management in smart grid: concepts, requirements and
  implementation.
\newblock {\em Journal of Big Data}, 4(1):1--19, 2017.

\bibitem{handschin1975bad}
Edmund Handschin, Fred~C Schweppe, Jurg Kohlas, and AAFA Fiechter.
\newblock Bad data analysis for power system state estimation.
\newblock {\em IEEE Transactions on Power Apparatus and Systems},
  94(2):329--337, 1975.

\bibitem{hauer2004performance}
John~F Hauer, Navin~B Bhatt, Kirit Shah, and Sharma Kolluri.
\newblock Performance of" wams east" in providing dynamic information for the
  north east blackout of august 14, 2003.
\newblock In {\em IEEE Power Engineering Society General Meeting, 2004.}, pages
  1685--1690. IEEE, 2004.

\bibitem{andersson2005causes}
G{\"o}ran Andersson, Peter Donalek, Richard Farmer, Nikos Hatziargyriou,
  Innocent Kamwa, Prabhashankar Kundur, Nelson Martins, John Paserba, Pouyan
  Pourbeik, Juan Sanchez-Gasca, et~al.
\newblock Causes of the 2003 major grid blackouts in north america and europe,
  and recommended means to improve system dynamic performance.
\newblock {\em IEEE transactions on Power Systems}, 20(4):1922--1928, 2005.

\bibitem{wu2005power}
Felix~F Wu, Khosrow Moslehi, and Anjan Bose.
\newblock Power system control centers: Past, present, and future.
\newblock {\em Proceedings of the IEEE}, 93(11):1890--1908, 2005.

\bibitem{phadke1993synchronized}
Arun~G Phadke.
\newblock Synchronized phasor measurements in power systems.
\newblock {\em IEEE Computer Applications in power}, 6(2):10--15, 1993.

\bibitem{li2008recognizing}
Dalu LI, Rui LI, Yuanzhang SUN, and Han CHEN.
\newblock Recognizing and correcting the wrong parameters in state estimation
  considering the wams measurements [j].
\newblock {\em Automation of Electric Power Systems}, 14, 2008.

\bibitem{anjia2005pmu}
Mao Anjia, Yu~Jiaxi, and Guo Zhizhong.
\newblock Pmu placement and data processing in wams that complements scada.
\newblock In {\em IEEE Power Engineering Society General Meeting, 2005}, pages
  780--783. IEEE, 2005.

\bibitem{morrow2012topology}
Kate~L Morrow, Erich Heine, Katherine~M Rogers, Rakesh~B Bobba, and Thomas~J
  Overbye.
\newblock Topology perturbation for detecting malicious data injection.
\newblock In {\em 2012 45th Hawaii International Conference on System
  Sciences}, pages 2104--2113. IEEE, 2012.

\bibitem{merrill1971bad}
Hyde~M Merrill and Fred~C Schweppe.
\newblock Bad data suppression in power system static state estimation.
\newblock {\em IEEE Transactions on Power Apparatus and Systems},
  (6):2718--2725, 1971.

\bibitem{kotiuga1982bad}
Willy~W Kotiuga and M~Vidyasagar.
\newblock Bad data rejection properties of weughted least absolute value
  techniques applied to static state estimation.
\newblock {\em IEEE Transactions on Power Apparatus and Systems}, (4):844--853,
  1982.

\bibitem{hongxun2018data}
Tian Hongxun, Wang Honggang, Zhou Kun, Shi Mingtai, Li~Haosong, Xu~Zhongping,
  Kang Taifeng, Li~Jin, and Cai Yaqi.
\newblock Data quality assessment for on-line monitoring and measuring system
  of power quality based on big data and data provenance theory.
\newblock In {\em 2018 IEEE 3rd International Conference on Cloud Computing and
  Big Data Analysis (ICCCBDA)}, pages 248--252. IEEE, 2018.

\bibitem{falcao1982power}
DM~Falcao, PA~Cooke, and A~Brameller.
\newblock Power system tracking state estimation and bad data processing.
\newblock {\em IEEE Transactions on Power Apparatus and Systems}, (2):325--333,
  1982.

\bibitem{monticelli1983reliable}
A~Monticelli and A~Garcia.
\newblock Reliable bad data processing for real-time state estimation.
\newblock {\em IEEE transactions on power apparatus and systems},
  (5):1126--1139, 1983.

\bibitem{peterson1988multiple}
William~L Peterson and AA~Girgis.
\newblock Multiple bad data detection in power system state estimation using
  linear programming.
\newblock In {\em The Twentieth Southeastern Symposium on System Theory}, pages
  405--406. IEEE Computer Society, 1988.

\bibitem{wu1988observability}
Felix~F Wu, W-HE Liu, and S-M Lun.
\newblock Observability analysis and bad data processing for state estimation
  with equality constraints.
\newblock {\em IEEE Transactions on Power Systems}, 3(2):541--548, 1988.

\bibitem{abur1990bad}
Ali Abur.
\newblock A bad data identification method for linear programming state
  estimation.
\newblock {\em IEEE Transactions on Power Systems}, 5(3):894--901, 1990.

\bibitem{huang2012state}
Yih-Fang Huang, Stefan Werner, Jing Huang, Neelabh Kashyap, and Vijay Gupta.
\newblock State estimation in electric power grids: Meeting new challenges
  presented by the requirements of the future grid.
\newblock {\em IEEE Signal Processing Magazine}, 29(5):33--43, 2012.

\bibitem{savulescu2009real}
Savu~Crivat Savulescu.
\newblock {\em Real-time stability assessment in modern power system control
  centers}, volume~42.
\newblock John Wiley \& Sons, 2009.

\bibitem{mazhabjafari2010feasibility}
A~Mazhabjafari, M~Kabiri, Kh~Khangholi, and S~Ghiami.
\newblock Feasibility study of running state estimation software in a large
  power system with a very weak rtu coverage condition.
\newblock In {\em 2010 Asia-Pacific Power and Energy Engineering Conference},
  pages 1--4. IEEE, 2010.

\bibitem{golden1989analytic}
Bruce~L Golden, Edward~A Wasil, and Patrick~T Harker.
\newblock The analytic hierarchy process.
\newblock {\em Applications and Studies, Berlin, Heidelberg}, 2, 1989.

\bibitem{saaty1989group}
Thomas~L Saaty.
\newblock Group decision making and the ahp.
\newblock In {\em The analytic hierarchy process}, pages 59--67. Springer,
  1989.

\end{thebibliography}

\end{document}